\documentclass[aps,twocolumn,showpacs,preprintnumbers,showkeys]{revtex4}

\usepackage[english]{babel}
\usepackage{dcolumn}
\usepackage{latexsym}
\usepackage{amsmath}
\usepackage{amssymb}
\usepackage{graphics}

\usepackage[breaklinks,pageanchor]{hyperref}
\usepackage{lipsum}
\usepackage{bookmark}
\usepackage{pdfpages}
\usepackage{fancyhdr}

\hypersetup{
  colorlinks=true,
  linkcolor=blue,
  filecolor=blue,
  urlcolor=blue,
  citecolor=blue,
}

\usepackage{float}
\usepackage{placeins}
\usepackage[active]{srcltx}
\usepackage{color,soul}

\begin{document}

\title{Unquenching the quark model in a non-perturbative scheme}

\author{Pablo G. Ortega}\email{pgortega@usal.es}
\author{David R. Entem}\email{entem@usal.es}
\author{Francisco Fern\'andez}\email{fdz@usal.es}
\affiliation{Grupo de F\'isica Nuclear and Instituto Universitario de
F\'isica Fundamental y Matem\'aticas (IUFFyM) \\ Universidad de Salamanca,
E-37008 Salamanca, Spain}

\date{\today}

\begin{abstract}
In recent years, the discovery in quarkonium spectrum of several states not predicted by the naive quark model has awakened a lot of interest. A possible description of such states requires the enlargement of the quark model by introducing quark-antiquark pair creation or continuum coupling effects. 
The unquenching of the quark models is a way to take these new components into account. 
In the spirit of the Cornell Model, this is usually done by coupling perturbatively a quark-antiquark state with definite quantum numbers to the meson-meson channel with closest threshold. 

In this work we present a method to coupled quark-antiquark states with meson-meson channels, taking into account effectively the 
non-perturbative coupling to all quark-antiquark states with the same quantum numbers.
The method will be applied to the study of the X(3872) resonance and a comparison with the perturbative calculation will be performed.
\end{abstract}


\maketitle

\section{Introduction}

Constituent quark models (CQM) have been extremely successful in describing the properties of hadrons 
such as the spectrum and the magnetic moments. However, since the earliest days of the hadron spectroscopy, it was 
realized~\cite{Eichten:1978tg} that such models neglect the contribution of higher Fock components (virtual~$q\bar q$
pairs, unitary loops or hadron-hadron channels) predicted by QCD.
Unquenching the quark model is a way to incorporate these components, in a similar way 
as unquenched lattice theories included dynamical quarks instead of static quarks (quenched theories).
It is worth to notice, though, that the name ``unquenched quark model'' refers to different approaches to include these higher Fock components.
Thus, Tornqvist and collaborators~\cite{Heikkila:1983wd,Ono:1983rd} use a 
``unitarized'' quark model to incorporate the effects of two meson channels to $c\bar c$ and $b\bar b$ spectrum. Van Beveren and Rupp~\cite{PhysRevD.21.772,PhysRevD.27.1527} showed the influence of continuum channels on the properties of hadrons, in a model which describes the meson as a system of one or more closed quark-antiquark states in interaction with several two meson channels. 
 More recently, Bijker and Santopinto~\cite{Bijker:2009up} developed an unquenched quark model for baryons, 
 in which a constituent quark model is modified to include, as a perturbation, a QCD-inspired quark-antiquark creation mechanism. 
 Under this assumption, the baryon wave function consists of a zeroth order three valence quark configuration plus a sum over meson-meson loops. 

The components beyond the naive constituent quark model became more relevant since 2003, when the $X(3872)$ and other XYZ states were discovered. At that time hadron spectroscopy begins to measure hadron states near two particle thresholds and in this situation loop corrections are relevant.

The $X(3872)$ resonance has been studied using different versions of the unquenched quark model in Ref.~\cite{PhysRevD.69.094019, Coito:2012vf, Ferretti:2013faa}. Eichten {\it et al.}~\cite{PhysRevD.69.094019} considered the influence of meson-meson channels using the Cornell coupled channels scheme~\cite{Eichten:1979ms}, obtaining a perturbative mass shift of the $c\bar c$ configuration 
below threshold and an additional decay width for configurations above threshold. Using the Resonance Spectrum expansion~\cite{vanBeveren:2003vs} Coito {\it et al.} showed that the $X(3872)$ is compatible with a description in terms of a regular $2^3P_1$ charmonium state with a renormalized mass, via opened and closed decay channels. Finally, in Ref.~\cite{Ferretti:2014xqa}, it was shown that one can describe the $X(3872)$ as a $\chi_{c1}(2P)$ interpreted as a $c\bar c$ core plus higher Fock components due to perturbative coupling to the meson-meson continuum.

A different point of view from the references mentioned above, where the goal is to study the influence of the two meson channels in the mass and width of the $q\bar q$ state, is presented in Ref.~\cite{Ortega:2010qq}. These authors start from a coupled channels approach for the $J^{PC}=1^{++}$ channel, to
study the influence of the $q\bar q$ channels in the meson-meson one, instead of the $q\bar q$ mass shift due to the coupling with the continuum. Solving the coupling for the $q\bar q$ states one arrives to a Schr\"odinger-type equation in which the meson-meson interaction gets modified by the coupling with the bare $q\bar q$ states, generating a resonance which can be identified with the $X(3872)$. Besides this state, one gets an orthogonal one which can be identified with the $X(3940)$. Both are different combinations of $q\bar q$ and meson-meson states.

A common characteristic of all these models is that they  have  to choose a particular~$q\bar q$ state which gets modified by the interaction with the two meson channels. Usually, the states ``closer'' to the threshold of the meson-meson channel are chosen, but sometimes it is difficult to decide which ones are relevant.  Moreover, this state is modified perturbatively, which is at least questionable when the $q\bar q$ states lie near in energy of the meson-meson threshold.

In an attempt to improve these caveats, we have developed a new scheme in which the contribution of all the $q\bar q$ states are automatically taken into account. The rationale of the method is to leave the $q\bar q$ radial wave function as unknown to be determined dynamically. Afterwards, the contribution of each 'quenched' state is determined by expanding the obtained wave function in a bare quark-antiquark base. In this way, one incorporates from scratch all possible $q\bar q $ states and also allows the deformation of the bare $q\bar q$ wave function due to interaction with the meson-meson channels.
Although the method is general for baryons and mesons, for the sake of clarity in this paper we will develop it only for mesons.

The paper is organized as follows. In Section~\ref{model} we first discuss the coupled channel formalism and the 
coupling mechanism between the different channels and the basic 
ingredients of the constituent quark model. Results and comments are 
presented in Section~\ref{results}. Finally, we summarize the main achievements of our 
calculation in Section~\ref{summary}.

\section{The Model}
\label{model}

\subsection{The constituent quark model}


The constituent quark model used in this work has been extensively
described elsewhere~\cite{Vijande:2004he,Segovia:2008zz} and therefore 
we will only summarize here its most
relevant aspects. The chiral symmetry of the original QCD Lagrangian appears
spontaneously broken in nature and, 
as a consequence, light quarks acquire a
dynamical mass. The simplest Lagrangian invariant under chiral rotations
must therefore contain chiral fields, and can be expressed as 
\begin{equation}
\label{lagrangian}
{\mathcal L}
=\overline{\psi }(i\, {\slash\!\!\! \partial} -M(q^{2})U^{\gamma_{5}})\,\psi 
\end{equation}
where  $U^{\gamma_5}=e^{i\frac{
\lambda _{a}}{f_{\pi }}\phi ^{a}\gamma _{5}}$ is 
the Goldstone boson fields matrix and $M(q^2)$ the dynamical
(constituent) mass. 
This Lagrangian has been derived in Ref.~\cite{Diakonov:2002fq}
as the low-energy limit in the instanton liquid model. In this model
the dynamical mass vanishes at large momenta and it is frozen at low 
momenta, for a value around 300 MeV. 
Similar results have also been obtained in lattice
calculations~\cite{Bowman:2005vx}. 
To simulate this behavior we parametrize 
the dynamical mass as
$M(q^{2})=m_{q}F(q^{2})$, 
where $m_{q}\simeq $ $300\,MeV$, and
\begin{equation}
F(q^{2})=\left[ \frac{\Lambda _{\chi}^{2}}{\Lambda _{\chi}^{2}+q^{2}}%
\right] ^{
{\frac12} 
} \, .
\end{equation}
The cut-off $\Lambda _{\chi}$ fixes the
chiral symmetry breaking scale.

The Goldstone boson field matrix $U^{\gamma_{5}}$ can be expanded in terms of 
boson fields,
\begin{equation}
U^{\gamma_{5}}=1+\frac{i}{f_{\pi }}\gamma^{5}\lambda^{a}\pi^{a}-\frac{1}{%
2f_{\pi}^{2}}\pi^{a}\pi^{a}+...
\end{equation}
The first term of the expansion generates the constituent quark mass while the
second gives rise to a one-boson exchange interaction between quarks. The
main contribution of the third term comes from the two-pion exchange which
has been simulated by means of a scalar exchange potential.

In the heavy quark sector 
chiral symmetry is explicitly broken and this type of interaction does not act. 
However it constrains the model parameters through the light meson 
phenomenology 
and provides a natural way to incorporate the pion exchange interaction in the 
open charm dynamics.

Below the chiral symmetry breaking scale quarks still interact
through gluon exchanges  described by
the Lagrangian
\begin{equation}
\label{Lg}
{\mathcal L}_{gqq}=
i\sqrt{4\pi \alpha _{s} }\,\,\overline{\psi }\gamma _{\mu }G^{\mu
}_c \lambda _{c}\psi  \, ,
\end{equation}
where $\lambda_{c}$ are the SU(3) color generators and $G^{\mu}_c$ the
gluon field. 
The other QCD nonperturbative effect corresponds to confinement,
which prevents from having colored hadrons.
Such a term can be physically interpreted in a picture in which
the quark and the antiquark are linked by a one-dimensional color flux-tube.
The spontaneous creation of light-quark pairs may
give rise at some scale to a breakup of the color flux-tube~\cite{Bali:2005fu}. This 
can be translated
into an screened potential~\cite{Born:1989iv} in such a way that the potential
saturates at some interquark distance
\begin{equation}
V_{CON}(\vec{r}_{ij})=\{-a_{c}\,(1-e^{-\mu_c\,r_{ij}})+ \Delta\}(\vec{%
\lambda^c}_{i}\cdot \vec{ \lambda^c}_{j}).
\end{equation}
Explicit 
expressions for these interactions and the value of the parameters are given in Ref.~\cite{Vijande:2004he,Segovia:2008zz}. 


\subsection{The unquenched meson spectrum}

Although over the time the procedure to incorporate new Fock components to the $q\bar q$ wave function has received several 
names (unitarized quark model~\cite{Heikkila:1983wd}, resonance spectrum expansion~\cite{vanBeveren:2003vs}, 
coupled channel formalism~\cite{Eichten:1978tg}) the basic idea behind the unquenched 
meson model is to assume that a hadron wave function with a fixed $J^P$ quantum numbers combines a zeroth order configuration plus a sum over the possible higher Fock components due to the creation of $q\bar q$ pairs

\begin{eqnarray}
|\Psi_A \rangle &=&\mathcal{N}| A \rangle + \sum_{BClj}\int d\vec{K} k^2dk |BClj,\vec{K},k \rangle \nonumber \\
&&\times \frac{\langle BClj,\vec{K},k|T^+|A\rangle}{E_0-E_B-E_C}
\label{ec:funondaunq}
\end{eqnarray}
where $T^+$ stand for the operator which couples the different components, usually the $^3P_0$ quark-antiquark pair creator, and $|A\rangle$ is an eigenstate of the bare hamiltonian $H_0|A\rangle=E_0|A\rangle$.

In practice what is done is to assume that the first Fock component, namely the $q\bar q$ structure, is renormalized via the influence of nearby meson-meson channels and, thus, a coupled channels calculation is performed including the bare $q\bar q$ state and the meson-meson channels. Solving the coupling with the 
meson-meson channels one obtains for the mass shift of the meson $|A\rangle$

\begin{equation}
\Sigma(E_A) = \sum_{BClj}\int dk^2dk\frac{|\langle BClj,\vec{K},k|T^+|A\rangle|^2}{E_A-E_{BC}}
\label{ec:massshift}
\end{equation}
where $E_{BC}$ is the kinetic energy of the meson-meson pair.

As stated in the introduction, this method has two important shortcomings. First of all, it focuses the study on the modification of the $q\bar q$ channel, avoiding the study of the meson-meson channel where interesting structures may appear. Second, it is chosen from the beginning one $q\bar q$ channel to be modified, neglecting those $q\bar q$ channels which may be generated in the interaction with the 
meson-meson channel.

For these reasons, we have developed a new scheme in which the contribution of all states are initially taken into account, being the dynamics the responsible of selecting the contribution of each bound state.

The hamiltonian we consider
\begin{equation}
H=H_0+V
\label{ec:ham}
\end{equation}
is the sum of an 'unperturbed' part $H_0$ and a second part $V$ which couples a $q\bar q$ system to a continuum made of meson-meson states.

Instead of expanding the wave function of the $q\bar q$ system in eigenstates of the $H_0$ hamiltonian and then solving the coupled channels equation with the meson-meson channels, we use a general wave function for the $q\bar q$ system to solve the coupled channels problem and then develop the solution of the $q\bar q$ system in the base of the bare $q\bar q$ states. In this way, the dynamics of the system is the element which determines the contribution of each bare state to the eigenstate, obtained from the two-body Schr\"odinger equation which includes the effect of the dynamics of the nearby meson-meson channels.

The meson wave functions to be used all along this work will be expressed
using the Gaussian Expansion Method~\cite{Hiyama:2003cu} (GEM), expanding the radial wave function in terms of basis
functions 

\begin{equation}\label{ec:GEMradial}
|\Phi_{\alpha}(r)\rangle=\sum_{n=1}^{n_{max}} c_{n}^\alpha |\phi^G_{nl}(r)\rangle
\end{equation} 
where $\alpha$ refers to the channel quantum numbers
and  $|\phi^G_{nl}(r)\rangle$ are Gaussian trial functions with
ranges in geometrical progression. This choice is useful for optimizing the ranges with a small
number of free parameters~\cite{Hiyama:2003cu}. In addition, the density  of the distribution of the 
Gaussian ranges in geometrical progression at small ranges is suitable for making the wave function 
correlate with short range potentials. 

To introduce higher Fock components in the $q\bar q$ wave function we assume that the hadronic state is
\begin{equation} 
\label{ec:funonda}
 | \Psi \rangle = | \Phi_{\alpha} \rangle
 + \sum_\beta \chi_\beta(P) |\phi_A \phi_B \beta \rangle
\end{equation}
where $| \Phi_{\alpha} \rangle$ is the $q\bar q$ wave function, 
$\phi_{M}$ are $q\bar q$  eigenstates describing 
the $A$ and $B$ mesons, 
$|\phi_A \phi_B \beta \rangle$ is the two meson state with $\beta$ quantum
numbers coupled to total $J^{PC}$ quantum numbers
and $\chi_\beta(P)$ is the relative wave 
function between the two mesons in the molecule. The meson-meson interaction will be derived from 
the $qq$ interaction using the Resonating Group Method (RGM)~\cite{Tang:1978zz}. 

We must notice that, although the Gaussian Expansion Method of the $q\bar q$ wave functions can be also used for the mesons constituting the molecular states, we will assume that the wave functions of these mesons are simple solutions of the Schr\"odinger two-body equation. 

In order to couple both sectors, we use the QCD-inspired  $^3P_0$ model~\cite{LeYaouanc:1972ae}, which gives a clear picture of the physical mechanism of the coupling. In this model, a quark pair is created from the vacuum with the vacuum quantum numbers. After
the pair creation, a recombination of the quark-antiquark of the initial meson with the $^3P_0$ pair follows to give the two final mesons.
The $^3P_0$ quark-antiquark pair-creation operator can be written as  
~\cite{Bonnaz:1999zj}
\begin{equation}
\begin{split}
\mathcal{T}=&-3\sqrt{2}\gamma'\sum_\mu \int d^3 p d^3p' \,\delta^{(3)}(p+p')\\
&\times \left[ \mathcal Y_1\left(\frac{p-p'}{2}\right) b_\mu^\dagger(p)
d_\nu^\dagger(p') \right]^{C=1,I=0,S=1,J=0}
\label{TBon}
\end{split}
\end{equation}
where $\mu$ ($\nu=\bar \mu$) are the quark (antiquark) quantum numbers and
$\gamma'=2^{5/2} \pi^{1/2}\gamma$ with $\gamma= \frac{g}{2m}$ is a 
dimensionless constant which gives the strength of 
the $q\bar q$ pair creation from the vacuum.

Finally, in the context of the $^3 P_0$ model, the transition operator $h_{\beta \alpha}(P)$ can be defined from the $\mathcal{T}$ operator as 
\begin{equation}
\label{Vab}
        \langle \phi_{M_1} \phi_{M_2} \beta | \mathcal{T}| \Phi_\alpha \rangle =
        P \, h_{\beta \alpha}(P) \,\delta^{(3)}(\vec P_{\mbox{cm}}) 
\end{equation}
where $P$ is the relative momentum of the two meson state. 

Now, we use Eq.~(\ref{ec:GEMradial}) to decompose $h_{\beta \alpha}(P)$ as
\begin{equation}\label{ec:3P0decomp}
        h_{\beta \alpha}(P) = \sum_{n=1}^{n_{max}} c_{n}^\alpha h_{\beta \alpha}^{n}(P) 
\end{equation}

Then, the coefficients $c_{n}^\alpha$ of the $q\bar q$ meson wave function and the eigenenergy $E$ are determined from the coupled-channel equations

\begin{widetext}
\begin{eqnarray}\label{ec:Ec-Res}
\sum_{\alpha,n} \ \left[\mathcal{H}_{n'n}^{\alpha'\alpha} - \mathcal{G}^{0\,\alpha'\alpha}_{n'n}(E)\right]c_{n}^{\alpha}& 
= & \sum_{n} EN_{n'n}^{\alpha'}c_{n}^{\alpha'}  \nonumber \\
 \sum_{\beta}\int H_{\beta'\beta}(P',P)\chi_{\beta}(P) P^2 dP + \sum_\alpha h_{\beta'\alpha}(P') & = & E \chi_{\beta'}(P')
\end{eqnarray}
\end{widetext}
where $N_{n'n}^{\alpha'}$ is the normalization matrix of the GEM trial Gaussian functions, 
$\mathcal{H}_{n'n}^{\alpha'\alpha}$ is the interquark Hamiltonian that defines the bare $q\bar q$ meson spectrum and 
$H_{\beta'\beta}$ is the RGM Hamiltonian for the two meson
states, obtained from the $q\bar q$ interaction. In the previous equations we have defined the perturbative mass shift 
$\mathcal{G}^{0\,\alpha'\alpha}_{n'n}$ as,
\begin{equation}
\mathcal{G}^{0\,\alpha'\alpha}_{n'n}(E) = -\delta^{\alpha'\alpha}\delta_{n'n}\sum_\beta \int h_{\alpha'\beta}^{n'}(P) \chi_\beta(P)P^2 dP,
\end{equation}
which codifies the coupling with molecular states. Apparently, the perturbative mass shift does not mix different $q\bar q$ channels, so
the only operator which mix them is the potential $V_{n'n}^{\alpha'\alpha}$ in $\mathcal{H}_{n'n}^{\alpha'\alpha}$. However, 
as will be detailed below, the molecular wave function
is dependent on different meson channels.

The previous coupled channel equations can be solved in a more elegant way through the $T$ matrix, solution of 
the Lippmann-Schwinger equation,
\begin{widetext}
\begin{equation}\label{ec:tmat}
T^{\beta'\beta}(P',P;E)=V^{\beta'\beta}(P', P;E)+\sum_{\beta''}\int 
V^{\beta'\beta''}(P',P'';E)
\frac{1}{E-E_{\beta''}(P'')}T^{\beta''\beta}(P'',P;E)
\,P''^2 dP'' 
\end{equation}
\end{widetext}
where $V^{\beta'\beta}(P',P)$ is the RGM potential. 

Using the $T$ matrix, we end up with a Schr\"odinger-like equation for the $c_{n}^\alpha$ coefficients,

\begin{eqnarray}\label{ec:Unquench}
\sum_{\alpha,n} \ \left[\mathcal{H}_{n'n}^{\alpha'\alpha} - \mathcal{G}^{\alpha'\alpha}_{n'n}(E)\right]c_{n}^{\alpha}& 
= & \sum_{n} EN_{n'n}^{\alpha'}c_{n}^{\alpha'} ,
\end{eqnarray}

In the previous equation we identify $\mathcal{G}^{\alpha'\alpha}_{n'n}$ as the energy-dependent complete mass-shift matrix

\begin{equation}
\mathcal{G}^{\alpha'\alpha}_{n'n}(E)=\sum_\beta \int dq q^2\frac{\bar h^{n'}_{\alpha'\beta}(q,E)h^n_{\beta \alpha}(q)}{q^2/2\mu-E-i0^+},
\end{equation}
where $\bar h^{\alpha' \beta}$ is the $^3P_0$ potential dressed by the RGM meson-meson interaction,

\begin{equation} \label{ec:VerticModf}
 \bar h^{\alpha \beta'} (p,E)=h_{\alpha \beta'}(p)-\sum_{\beta}\int d^3q \frac{T^{\beta'\beta}(p,q,E)h_{\alpha \beta}(q)}{q^2/2\mu-E-i0^+},\\
\end{equation}
which can be decomposed in the GEM basis as $h^{\alpha\beta}$ in Eq.~(\ref{ec:3P0decomp}).

In order to find molecular states above and below thresholds in the same 
formalism we have to analytically continue all the potentials for complex momenta.
Therefore, resonances are solutions of the Eq.~(\ref{ec:Unquench}), with the pole position
solved by the Broyden method~\cite{Broyden:1965}.

The molecular wave function is related with the $c_{n}^\alpha$ coeficients of the meson $q\bar q$ state,

\begin{equation} \label{ec:FunMTexac}
 \chi_\beta(p) =-\sum_{\alpha,n}\frac{\bar h^n_{\alpha \beta}(p,E)c_n^\alpha(E)}{p^2/2\mu-E -i0^+}
\end{equation}
so the global normalization of the $q\bar q$-$qq\bar q\bar q$ mix state is 

\begin{equation}
 1 = \sum_{\alpha,\alpha',n,n'} c_{n'}^{\alpha'}\ N_{n'n}^{\alpha'\alpha}\ c_n^\alpha + \langle \chi_\beta | \chi_\beta \rangle
\end{equation}

The partial decay widths can be defined through the complete S-matrix of the mix channel, as detailed in~\cite{Ortega:2012rs}.

\section{Calculations, results and discussions}
\label{results}

In order to compare the results of the proposed scheme with the perturbative one we perform a similar calculation as in Ref.~\cite{Ortega:2010qq},
namely, a coupled channel calculation including the $1^{++}$ $q\bar q$ sector and the 
$D^0 \bar D^{0*}$ and $D^{\pm}D^{*\mp}$ channels.
We, first of all, perform a calculation with the same parameters as~\cite{Ortega:2010qq} in the charge basis

\begin{equation}
\label{ec:charb1}
|D^{\pm}D^{*\mp}\rangle =\frac{1}{\sqrt{2}}\left(|D\bar D^*I=0\rangle-|D\bar D^*I=1\rangle \right)
\end{equation}
 
\begin{equation}
\label{ec:charb2}
|D^{0}\bar D^{*0}\rangle =\frac{1}{\sqrt{2}}\left(|D\bar D^*I=0\rangle+|D\bar D^*I=1\rangle \right)
\end{equation}

Isospin symmetry is explicitly broken taking the experimental threshold difference into account in our equations and solving for the charged and the neutral components.

One can see in Table~\ref{tab:maschp} that when we use the value of $\gamma= 0.26$ we get three states with energies close to the $X(3940)$, the $X(3872)$ and the $c\bar c$ $M=3510$ MeV$/c^2$. The first two states are mixtures of $c\bar c$ and $D\bar D^*$ components, being the $X(3940)$ predominantly $c\bar c$ and the $X(3872)$ mostly $D\bar D^*$. The third state is clearly a $q\bar q$ state. If we project the $q\bar q$ component in the base of bare $c\bar c$ states (Table~\ref{tab:ccbar decomp}) we realize that this third state is an almost pure $1^3P_1$ $c\bar c$ state whereas the other two are predominantly $2^3P_1$. 

One can analyze the isospin content of 
the $D\bar D^*$ component of the different states. The only one which shows a sizeable $I=1$ component is the one we have identified with the $X(3872)$. The other two are basically $I=0$ states. Notice that all the interactions are isospin conserving interactions and the reason to have a non-zero $I=1$ component
for the $X(3872)$ is that the binding energy of the $D\bar D^*$ component for this state is smaller than the isospin breaking of the $D^{(*)0}$
and $D^{(*)\pm}$ masses which enhances the $D^0\bar D^{*0}$ component at large distances.

As in Ref.~\cite{Ortega:2010qq}, we have fine-tuned the $^3P_0$ parameter to get the right binding energy of the $X(3872)$. The results are shown in the second part of Table~\ref{tab:maschp}. We get again three states. The first one, with a mass of $M=3943.9$ Mev$/c^2$, can be identified with the 
$X(3940)$~\cite{PDG2014} and is a mixture of $57\%$ of $q\bar q$ and $43\%$ of $D\bar D^*$ molecule with isospin $I=0$. The $q\bar q$ component is basically a $2^3P_1$ state. The second state has now the right energy of the $X(3872)$ resonance. The $D^0\bar D^{*0}$ clearly dominates its structure with a $94\%$ probability, giving a $55\%$ probability for the isospin $0$ component and $39\%$ for isospin $1$, which is enhanced in this case because the binding energy is much smaller compare with mass difference of the neutral and charged $D\bar D^*$ components. As the isospin breaking is a threshold effect, it grows as we get closer to the $X(3872)$ physical mass. The $q\bar q$ component is, as in the previous state, a $2^3P_1$ bare state. Finally we get an almost pure $I=0$ $1^3P_1$ state with a mass of $M=3480.75$ Mev$/c^2$.

\begin{table}[!t]
\begin{center}
\begin{tabular}{ccccccc}
\hline
\hline
$\gamma ^3P_0$ & $M(MeV)$ & $c\bar c$  & $D^0\bar D^{*0}$ & $D^{\pm}D^{*\pm}$ & I=0      & I=1 \\
\hline
               & $3948.89$   &  $56.71\% $ & $22.47\%$  & $20.82\%$       & $43.10\%$ & $0.25\%$ \\
0.260          & $3867.36$   &$30.22\%$    & $51.37\%$  & $18.40\%$       & $64.72\%$ &  $5.06\%$ \\
		 & $3468.29$   & $95.70\%$   & $2.18\%$  & $2.12\%$       & $4.30\%$ &  $0.0\%$ \\
\hline
              & 3944.58   & $57.03\%$   & $22.07\%$  &  $20.89\%$      & $42.72\%$ & $0.43\%$ \\
0.218         & 3871.76   & $3.62\%$   &$93.99\%$   & $2.39\%$       & $54.98\%$ & $39.34\%$ \\
              & 3478.55   & $96.84\%$   & $1.60\%$  & $1.56\%$       & $3.16\%$ & $0.0\%$ \\ 

\hline
\hline
\end{tabular}
\caption{\label{tab:maschp} Mass and channel probabilities for the three components in two different calculations.}
\end{center}
\end{table}

\begin{table}[!t]
\begin{center}
\begin{tabular}{ccccccc}
\hline
\hline
$\gamma ^3P_0$ & $M(MeV)$ & $c\bar c$  & $1^3P_1$ & $2^3P_1$ & $3^3P_1$ & $4^3P_1$ \\
\hline
               & $3948.89$   &  $56.71\% $ & $1.61\%$  & $96.33\%$  & $1.28\%$ & $0.78\%$ \\
0.260           & $3867.36$   &$30.22\%$    & $1.80\%$  & $98.14\%$       & $0.06\%$ &  $0.0\%$ \\
		 & $3468.29$   & $95.70\%$   & $99.99\%$  & $0.01\%$       & $0.0\%$ &  $0.0\%$ \\
\hline
              & 3944.58   & $57.03\%$   & $0.23\%$  &  $99.49\%$      & $0.28\%$ & $0.0\%$ \\
0.218         & 3871.76   & $3.62\%$   &$2.11\%$   & $97.75\%$       & $0.14\%$ & $0.0\%$ \\
              & 3478.55   & $96.84\%$   & $100.0\%$  & $0.0\%$       & $0.0\%$ & $0.0\%$ \\ 

\hline
\hline
\end{tabular}
\caption{\label{tab:ccbar decomp} Decomposition of the $c\bar c$ component in bare $c\bar c$ states.}
\end{center}
\end{table}

The scenario drawn by these results consist on two bare states (a molecule and a $2^3P_1$ $q\bar q$ state), which mix together to give the two physical states, and a third state which participate slightly into the game, although as we will see, its contribution is important.
The result is similar to those of Ref.~\cite{Ortega:2010qq} but now it appear dynamically without the need to make educated guesses about the bare states involved in the calculation. In this way we are confident that all the physics is included into the model.

Although not addressed in Ref.~\cite{Ortega:2010qq}, we have calculated with our wave functions the two decay rates that represent a challenge in the description of the $X(3782)$ structure, namely
\begin{equation}
\label{ec:ratioI}
R_I=\frac{\mathcal{B}\left(X(3872)\rightarrow \omega J/\psi\right)}{\mathcal{B}\left(X(3872)\rightarrow \pi^-\pi^+ J/\psi\right)}
\end{equation}

\begin{equation}
\label{ec:ratioI}
R_{\gamma}=\frac{\mathcal{B}\left(X(3872)\rightarrow \psi (2S)\gamma\right)}{\mathcal{B}\left(X(3872)\rightarrow J/\psi \gamma\right)}
\end{equation}
$R_I$ has been measured by BaBar Collaboration~\cite{PhysRevD.82.011101} obtaining $R_I=0.8\pm 0.3$. A value of 
$R_\gamma=2.46\pm 0.64\pm0.29$ has been recently reported by  LHCb Collaboration~\cite{Aaij2014665}. 

Whereas $R_I$, which shown a large isospin symmetry breaking, has been used to justify that the $X(3872)$ should be an hadronic molecule, the high value of $R_{\gamma} $ was interpreted as a strong evidence that the $X(3872)$ cannot be a pure hadronic molecule, based on the claim that the contributions of the molecular component to this ratio should be small~\cite{Swanson2004197}.

Our $X(3872)$ wave function allows for a simultaneous description of both ratios. 
To calculate $R_{\gamma} $ we have assumed that the decay proceed through the $q\bar q$ component, obtaining the value $R_{\gamma}=0.39$
which is clearly far from the experimental value. A similar result is obtained in~\cite{Takizawa:2014nma}, where it is shown that if the $1^3P_1$ states is neglected the ratio approaches to the experimental value. Then, it seems that either the two meson components have to be taken into account or a more careful calculation of this decay should be done.

The obtained value for the first ratio is $R_I=1.5$ that is close to the experimental result.

\section{Summary}
\label{summary}

In this work we present a method to coupled quark-antiquark states with meson-meson channels, taking into account effectively the 
non-perturbative coupling to all quark-antiquark states with the same quantum numbers. Instead of expanding the wave function of the $q\bar q$ system in eigenstates of the $H_0$ hamiltonian and then solving the coupled channels equation with the meson-meson channels, we use a general wave function for the $q\bar q$ system to solve the coupled channels problem and then develop the solution of the $q\bar q$ system in the base of the bare $q\bar q$ states. 
The method is applied to the coupling of the $X(3872)$ resonance to the $1^{++}$ $q\bar q$ states and the results compared with those of the perturbative calculation of Ref.~\cite{Ortega:2010qq}.

\begin{acknowledgments}
This work has been funded by Ministerio de Econom\'ia, Industria y
Competitividad under Contract No. FPA2016-77177-C2-2-P.
P.G.O. acknowledges the financial support from Spanish MINECO's Juan de la Cierva-Incorporaci\'on programme, Grant Agreement No. IJCI-2016-28525.
\end{acknowledgments}

\bibliographystyle{apsrev4-1}
\bibliography{AHEPEntem}

\begin{thebibliography}{30}%
\makeatletter
\providecommand \@ifxundefined [1]{%
 \@ifx{#1\undefined}
}%
\providecommand \@ifnum [1]{%
 \ifnum #1\expandafter \@firstoftwo
 \else \expandafter \@secondoftwo
 \fi
}%
\providecommand \@ifx [1]{%
 \ifx #1\expandafter \@firstoftwo
 \else \expandafter \@secondoftwo
 \fi
}%
\providecommand \natexlab [1]{#1}%
\providecommand \enquote  [1]{``#1''}%
\providecommand \bibnamefont  [1]{#1}%
\providecommand \bibfnamefont [1]{#1}%
\providecommand \citenamefont [1]{#1}%
\providecommand \href@noop [0]{\@secondoftwo}%
\providecommand \href [0]{\begingroup \@sanitize@url \@href}%
\providecommand \@href[1]{\@@startlink{#1}\@@href}%
\providecommand \@@href[1]{\endgroup#1\@@endlink}%
\providecommand \@sanitize@url [0]{\catcode `\\12\catcode `\$12\catcode
  `\&12\catcode `\#12\catcode `\^12\catcode `\_12\catcode `\%12\relax}%
\providecommand \@@startlink[1]{}%
\providecommand \@@endlink[0]{}%
\providecommand \url  [0]{\begingroup\@sanitize@url \@url }%
\providecommand \@url [1]{\endgroup\@href {#1}{\urlprefix }}%
\providecommand \urlprefix  [0]{URL }%
\providecommand \Eprint [0]{\href }%
\providecommand \doibase [0]{http://dx.doi.org/}%
\providecommand \selectlanguage [0]{\@gobble}%
\providecommand \bibinfo  [0]{\@secondoftwo}%
\providecommand \bibfield  [0]{\@secondoftwo}%
\providecommand \translation [1]{[#1]}%
\providecommand \BibitemOpen [0]{}%
\providecommand \bibitemStop [0]{}%
\providecommand \bibitemNoStop [0]{.\EOS\space}%
\providecommand \EOS [0]{\spacefactor3000\relax}%
\providecommand \BibitemShut  [1]{\csname bibitem#1\endcsname}%
\let\auto@bib@innerbib\@empty
\bibitem [{\citenamefont {Eichten}\ \emph {et~al.}()\citenamefont {Eichten},
  \citenamefont {Gottfried}, \citenamefont {Kinoshita}, \citenamefont {Lane},\
  and\ \citenamefont {Yan}}]{Eichten:1978tg}%
  \BibitemOpen
  \bibfield  {author} {\bibinfo {author} {\bibfnamefont {E.}~\bibnamefont
  {Eichten}}, \bibinfo {author} {\bibfnamefont {K.}~\bibnamefont {Gottfried}},
  \bibinfo {author} {\bibfnamefont {T.}~\bibnamefont {Kinoshita}}, \bibinfo
  {author} {\bibfnamefont {K.}~\bibnamefont {Lane}}, \ and\ \bibinfo {author}
  {\bibfnamefont {T.-M.}\ \bibnamefont {Yan}},\ }\href {\doibase
  10.1103/PhysRevD.17.3090} {\bibfield  {journal} {\bibinfo  {journal} {Phys.
  Rev.}\ }\textbf {\bibinfo {volume} {D17}},\ \bibinfo {pages}
  {3090}}\BibitemShut {NoStop}%
\bibitem [{\citenamefont {Heikkila}\ \emph {et~al.}(1984)\citenamefont
  {Heikkila}, \citenamefont {Ono},\ and\ \citenamefont
  {Tornqvist}}]{Heikkila:1983wd}%
  \BibitemOpen
  \bibfield  {author} {\bibinfo {author} {\bibfnamefont {K.}~\bibnamefont
  {Heikkila}}, \bibinfo {author} {\bibfnamefont {S.}~\bibnamefont {Ono}}, \
  and\ \bibinfo {author} {\bibfnamefont {N.~A.}\ \bibnamefont {Tornqvist}},\
  }\href {\doibase 10.1103/PhysRevD.29.110} {\bibfield  {journal} {\bibinfo
  {journal} {Phys. Rev.}\ }\textbf {\bibinfo {volume} {D29}},\ \bibinfo {pages}
  {110} (\bibinfo {year} {1984})},\ \bibinfo {note} {[Erratum: Phys.
  Rev.D29,2136(1984)]}\BibitemShut {NoStop}%
\bibitem [{\citenamefont {Ono}\ and\ \citenamefont
  {Tornqvist}(1984)}]{Ono:1983rd}%
  \BibitemOpen
  \bibfield  {author} {\bibinfo {author} {\bibfnamefont {S.}~\bibnamefont
  {Ono}}\ and\ \bibinfo {author} {\bibfnamefont {N.~A.}\ \bibnamefont
  {Tornqvist}},\ }\href {\doibase 10.1007/BF01558041} {\bibfield  {journal}
  {\bibinfo  {journal} {Z. Phys.}\ }\textbf {\bibinfo {volume} {C23}},\
  \bibinfo {pages} {59} (\bibinfo {year} {1984})}\BibitemShut {NoStop}%
\bibitem [{\citenamefont {van Beveren}\ \emph {et~al.}(1980)\citenamefont {van
  Beveren}, \citenamefont {Dullemond},\ and\ \citenamefont
  {Rupp}}]{PhysRevD.21.772}%
  \BibitemOpen
  \bibfield  {author} {\bibinfo {author} {\bibfnamefont {E.}~\bibnamefont {van
  Beveren}}, \bibinfo {author} {\bibfnamefont {C.}~\bibnamefont {Dullemond}}, \
  and\ \bibinfo {author} {\bibfnamefont {G.}~\bibnamefont {Rupp}},\ }\href
  {\doibase 10.1103/PhysRevD.21.772} {\bibfield  {journal} {\bibinfo  {journal}
  {Phys. Rev. D}\ }\textbf {\bibinfo {volume} {21}},\ \bibinfo {pages} {772}
  (\bibinfo {year} {1980})}\BibitemShut {NoStop}%
\bibitem [{\citenamefont {van Beveren}\ \emph {et~al.}(1983)\citenamefont {van
  Beveren}, \citenamefont {Rupp}, \citenamefont {Rijken},\ and\ \citenamefont
  {Dullemond}}]{PhysRevD.27.1527}%
  \BibitemOpen
  \bibfield  {author} {\bibinfo {author} {\bibfnamefont {E.}~\bibnamefont {van
  Beveren}}, \bibinfo {author} {\bibfnamefont {G.}~\bibnamefont {Rupp}},
  \bibinfo {author} {\bibfnamefont {T.~A.}\ \bibnamefont {Rijken}}, \ and\
  \bibinfo {author} {\bibfnamefont {C.}~\bibnamefont {Dullemond}},\ }\href
  {\doibase 10.1103/PhysRevD.27.1527} {\bibfield  {journal} {\bibinfo
  {journal} {Phys. Rev. D}\ }\textbf {\bibinfo {volume} {27}},\ \bibinfo
  {pages} {1527} (\bibinfo {year} {1983})}\BibitemShut {NoStop}%
\bibitem [{\citenamefont {Bijker}\ \emph {et~al.}(2009)\citenamefont {Bijker},
  \citenamefont {Santopinto},\ and\ \citenamefont
  {Santopinto}}]{Bijker:2009up}%
  \BibitemOpen
  \bibfield  {author} {\bibinfo {author} {\bibfnamefont {R.}~\bibnamefont
  {Bijker}}, \bibinfo {author} {\bibfnamefont {E.}~\bibnamefont {Santopinto}},
  \ and\ \bibinfo {author} {\bibfnamefont {E.}~\bibnamefont {Santopinto}},\
  }\href {\doibase 10.1103/PhysRevC.80.065210} {\bibfield  {journal} {\bibinfo
  {journal} {Phys. Rev.}\ }\textbf {\bibinfo {volume} {C80}},\ \bibinfo {pages}
  {065210} (\bibinfo {year} {2009})},\ \Eprint {http://arxiv.org/abs/0912.4494}
  {arXiv:0912.4494 [nucl-th]} \BibitemShut {NoStop}%
\bibitem [{\citenamefont {Eichten}\ \emph {et~al.}(2004)\citenamefont
  {Eichten}, \citenamefont {Lane},\ and\ \citenamefont
  {Quigg}}]{PhysRevD.69.094019}%
  \BibitemOpen
  \bibfield  {author} {\bibinfo {author} {\bibfnamefont {E.~J.}\ \bibnamefont
  {Eichten}}, \bibinfo {author} {\bibfnamefont {K.}~\bibnamefont {Lane}}, \
  and\ \bibinfo {author} {\bibfnamefont {C.}~\bibnamefont {Quigg}},\ }\href
  {\doibase 10.1103/PhysRevD.69.094019} {\bibfield  {journal} {\bibinfo
  {journal} {Phys. Rev. D}\ }\textbf {\bibinfo {volume} {69}},\ \bibinfo
  {pages} {094019} (\bibinfo {year} {2004})}\BibitemShut {NoStop}%
\bibitem [{\citenamefont {Coito}\ \emph {et~al.}(2013)\citenamefont {Coito},
  \citenamefont {Rupp},\ and\ \citenamefont {van Beveren}}]{Coito:2012vf}%
  \BibitemOpen
  \bibfield  {author} {\bibinfo {author} {\bibfnamefont {S.}~\bibnamefont
  {Coito}}, \bibinfo {author} {\bibfnamefont {G.}~\bibnamefont {Rupp}}, \ and\
  \bibinfo {author} {\bibfnamefont {E.}~\bibnamefont {van Beveren}},\ }\href
  {\doibase 10.1140/epjc/s10052-013-2351-8} {\bibfield  {journal} {\bibinfo
  {journal} {Eur. Phys. J.}\ }\textbf {\bibinfo {volume} {C73}},\ \bibinfo
  {pages} {2351} (\bibinfo {year} {2013})},\ \Eprint
  {http://arxiv.org/abs/1212.0648} {arXiv:1212.0648 [hep-ph]} \BibitemShut
  {NoStop}%
\bibitem [{\citenamefont {Ferretti}\ \emph {et~al.}(2013)\citenamefont
  {Ferretti}, \citenamefont {Galatà},\ and\ \citenamefont
  {Santopinto}}]{Ferretti:2013faa}%
  \BibitemOpen
  \bibfield  {author} {\bibinfo {author} {\bibfnamefont {J.}~\bibnamefont
  {Ferretti}}, \bibinfo {author} {\bibfnamefont {G.}~\bibnamefont {Galatà}}, \
  and\ \bibinfo {author} {\bibfnamefont {E.}~\bibnamefont {Santopinto}},\
  }\href {\doibase 10.1103/PhysRevC.88.015207} {\bibfield  {journal} {\bibinfo
  {journal} {Phys. Rev.}\ }\textbf {\bibinfo {volume} {C88}},\ \bibinfo {pages}
  {015207} (\bibinfo {year} {2013})},\ \Eprint {http://arxiv.org/abs/1302.6857}
  {arXiv:1302.6857 [hep-ph]} \BibitemShut {NoStop}%
\bibitem [{\citenamefont {Eichten}\ \emph {et~al.}(1980)\citenamefont
  {Eichten}, \citenamefont {Gottfried}, \citenamefont {Kinoshita},
  \citenamefont {Lane},\ and\ \citenamefont {Yan}}]{Eichten:1979ms}%
  \BibitemOpen
  \bibfield  {author} {\bibinfo {author} {\bibfnamefont {E.}~\bibnamefont
  {Eichten}}, \bibinfo {author} {\bibfnamefont {K.}~\bibnamefont {Gottfried}},
  \bibinfo {author} {\bibfnamefont {T.}~\bibnamefont {Kinoshita}}, \bibinfo
  {author} {\bibfnamefont {K.}~\bibnamefont {Lane}}, \ and\ \bibinfo {author}
  {\bibfnamefont {T.-M.}\ \bibnamefont {Yan}},\ }\href {\doibase
  10.1103/PhysRevD.21.203} {\bibfield  {journal} {\bibinfo  {journal} {Phys.
  Rev.}\ }\textbf {\bibinfo {volume} {D21}},\ \bibinfo {pages} {203} (\bibinfo
  {year} {1980})}\BibitemShut {NoStop}%
\bibitem [{\citenamefont {van Beveren}\ and\ \citenamefont
  {Rupp}(2006)}]{vanBeveren:2003vs}%
  \BibitemOpen
  \bibfield  {author} {\bibinfo {author} {\bibfnamefont {E.}~\bibnamefont {van
  Beveren}}\ and\ \bibinfo {author} {\bibfnamefont {G.}~\bibnamefont {Rupp}},\
  }\href@noop {} {\bibfield  {journal} {\bibinfo  {journal} {Int. J. Theor.
  Phys. Group Theor. Nonlin. Opt.}\ }\textbf {\bibinfo {volume} {11}},\
  \bibinfo {pages} {179} (\bibinfo {year} {2006})},\ \Eprint
  {http://arxiv.org/abs/hep-ph/0304105} {arXiv:hep-ph/0304105 [hep-ph]}
  \BibitemShut {NoStop}%
\bibitem [{\citenamefont {Ferretti}\ \emph {et~al.}(2014)\citenamefont
  {Ferretti}, \citenamefont {Galatà},\ and\ \citenamefont
  {Santopinto}}]{Ferretti:2014xqa}%
  \BibitemOpen
  \bibfield  {author} {\bibinfo {author} {\bibfnamefont {J.}~\bibnamefont
  {Ferretti}}, \bibinfo {author} {\bibfnamefont {G.}~\bibnamefont {Galatà}}, \
  and\ \bibinfo {author} {\bibfnamefont {E.}~\bibnamefont {Santopinto}},\
  }\href {\doibase 10.1103/PhysRevD.90.054010} {\bibfield  {journal} {\bibinfo
  {journal} {Phys. Rev.}\ }\textbf {\bibinfo {volume} {D90}},\ \bibinfo {pages}
  {054010} (\bibinfo {year} {2014})},\ \Eprint {http://arxiv.org/abs/1401.4431}
  {arXiv:1401.4431 [nucl-th]} \BibitemShut {NoStop}%
\bibitem [{\citenamefont {Ortega}\ \emph {et~al.}(2010)\citenamefont {Ortega},
  \citenamefont {Segovia}, \citenamefont {Entem},\ and\ \citenamefont
  {Fernandez}}]{Ortega:2010qq}%
  \BibitemOpen
  \bibfield  {author} {\bibinfo {author} {\bibfnamefont {P.}~\bibnamefont
  {Ortega}}, \bibinfo {author} {\bibfnamefont {J.}~\bibnamefont {Segovia}},
  \bibinfo {author} {\bibfnamefont {D.}~\bibnamefont {Entem}}, \ and\ \bibinfo
  {author} {\bibfnamefont {F.}~\bibnamefont {Fernandez}},\ }\href {\doibase
  10.1103/PhysRevD.81.054023} {\bibfield  {journal} {\bibinfo  {journal} {Phys.
  Rev.}\ }\textbf {\bibinfo {volume} {D81}},\ \bibinfo {pages} {054023}
  (\bibinfo {year} {2010})}\BibitemShut {NoStop}%
\bibitem [{\citenamefont {Vijande}\ \emph {et~al.}(2005)\citenamefont
  {Vijande}, \citenamefont {Fernandez},\ and\ \citenamefont
  {Valcarce}}]{Vijande:2004he}%
  \BibitemOpen
  \bibfield  {author} {\bibinfo {author} {\bibfnamefont {J.}~\bibnamefont
  {Vijande}}, \bibinfo {author} {\bibfnamefont {F.}~\bibnamefont {Fernandez}},
  \ and\ \bibinfo {author} {\bibfnamefont {A.}~\bibnamefont {Valcarce}},\
  }\href {\doibase 10.1088/0954-3899/31/5/017} {\bibfield  {journal} {\bibinfo
  {journal} {J. Phys.}\ }\textbf {\bibinfo {volume} {G31}},\ \bibinfo {pages}
  {481} (\bibinfo {year} {2005})}\BibitemShut {NoStop}%
\bibitem [{\citenamefont {Segovia}\ \emph {et~al.}(2008)\citenamefont
  {Segovia}, \citenamefont {Yasser}, \citenamefont {Entem},\ and\ \citenamefont
  {Fernandez}}]{Segovia:2008zz}%
  \BibitemOpen
  \bibfield  {author} {\bibinfo {author} {\bibfnamefont {J.}~\bibnamefont
  {Segovia}}, \bibinfo {author} {\bibfnamefont {A.}~\bibnamefont {Yasser}},
  \bibinfo {author} {\bibfnamefont {D.}~\bibnamefont {Entem}}, \ and\ \bibinfo
  {author} {\bibfnamefont {F.}~\bibnamefont {Fernandez}},\ }\href {\doibase
  10.1103/PhysRevD.78.114033} {\bibfield  {journal} {\bibinfo  {journal} {Phys.
  Rev.}\ }\textbf {\bibinfo {volume} {D78}},\ \bibinfo {pages} {114033}
  (\bibinfo {year} {2008})}\BibitemShut {NoStop}%
\bibitem [{\citenamefont {Diakonov}(2003)}]{Diakonov:2002fq}%
  \BibitemOpen
  \bibfield  {author} {\bibinfo {author} {\bibfnamefont {D.}~\bibnamefont
  {Diakonov}},\ }\href {\doibase 10.1016/S0146-6410(03)90014-7} {\bibfield
  {journal} {\bibinfo  {journal} {Prog. Part. Nucl. Phys.}\ }\textbf {\bibinfo
  {volume} {51}},\ \bibinfo {pages} {173} (\bibinfo {year} {2003})},\ \Eprint
  {http://arxiv.org/abs/hep-ph/0212026} {arXiv:hep-ph/0212026 [hep-ph]}
  \BibitemShut {NoStop}%
\bibitem [{\citenamefont {Bowman}\ \emph {et~al.}(2005)\citenamefont {Bowman},
  \citenamefont {Heller}, \citenamefont {Leinweber}, \citenamefont
  {Parappilly}, \citenamefont {Williams},\ and\ \citenamefont
  {Zhang}}]{Bowman:2005vx}%
  \BibitemOpen
  \bibfield  {author} {\bibinfo {author} {\bibfnamefont {P.~O.}\ \bibnamefont
  {Bowman}}, \bibinfo {author} {\bibfnamefont {U.~M.}\ \bibnamefont {Heller}},
  \bibinfo {author} {\bibfnamefont {D.~B.}\ \bibnamefont {Leinweber}}, \bibinfo
  {author} {\bibfnamefont {M.~B.}\ \bibnamefont {Parappilly}}, \bibinfo
  {author} {\bibfnamefont {A.~G.}\ \bibnamefont {Williams}}, \ and\ \bibinfo
  {author} {\bibfnamefont {J.-b.}\ \bibnamefont {Zhang}},\ }\href {\doibase
  10.1103/PhysRevD.71.054507} {\bibfield  {journal} {\bibinfo  {journal} {Phys.
  Rev.}\ }\textbf {\bibinfo {volume} {D71}},\ \bibinfo {pages} {054507}
  (\bibinfo {year} {2005})},\ \Eprint {http://arxiv.org/abs/hep-lat/0501019}
  {arXiv:hep-lat/0501019 [hep-lat]} \BibitemShut {NoStop}%
\bibitem [{\citenamefont {Bali}\ \emph {et~al.}(2005)\citenamefont {Bali},
  \citenamefont {Neff}, \citenamefont {Duessel}, \citenamefont {Lippert},\ and\
  \citenamefont {Schilling}}]{Bali:2005fu}%
  \BibitemOpen
  \bibfield  {author} {\bibinfo {author} {\bibfnamefont {G.~S.}\ \bibnamefont
  {Bali}}, \bibinfo {author} {\bibfnamefont {H.}~\bibnamefont {Neff}}, \bibinfo
  {author} {\bibfnamefont {T.}~\bibnamefont {Duessel}}, \bibinfo {author}
  {\bibfnamefont {T.}~\bibnamefont {Lippert}}, \ and\ \bibinfo {author}
  {\bibfnamefont {K.}~\bibnamefont {Schilling}} (\bibinfo {collaboration}
  {SESAM}),\ }\href {\doibase 10.1103/PhysRevD.71.114513} {\bibfield  {journal}
  {\bibinfo  {journal} {Phys. Rev.}\ }\textbf {\bibinfo {volume} {D71}},\
  \bibinfo {pages} {114513} (\bibinfo {year} {2005})},\ \Eprint
  {http://arxiv.org/abs/hep-lat/0505012} {arXiv:hep-lat/0505012 [hep-lat]}
  \BibitemShut {NoStop}%
\bibitem [{\citenamefont {Born}\ \emph {et~al.}(1989)\citenamefont {Born},
  \citenamefont {Laermann}, \citenamefont {Pirch}, \citenamefont {Walsh},\ and\
  \citenamefont {Zerwas}}]{Born:1989iv}%
  \BibitemOpen
  \bibfield  {author} {\bibinfo {author} {\bibfnamefont {K.~D.}\ \bibnamefont
  {Born}}, \bibinfo {author} {\bibfnamefont {E.}~\bibnamefont {Laermann}},
  \bibinfo {author} {\bibfnamefont {N.}~\bibnamefont {Pirch}}, \bibinfo
  {author} {\bibfnamefont {T.~F.}\ \bibnamefont {Walsh}}, \ and\ \bibinfo
  {author} {\bibfnamefont {P.~M.}\ \bibnamefont {Zerwas}},\ }\href {\doibase
  10.1103/PhysRevD.40.1653} {\bibfield  {journal} {\bibinfo  {journal} {Phys.
  Rev.}\ }\textbf {\bibinfo {volume} {D40}},\ \bibinfo {pages} {1653} (\bibinfo
  {year} {1989})}\BibitemShut {NoStop}%
\bibitem [{\citenamefont {Hiyama}\ \emph {et~al.}(2003)\citenamefont {Hiyama},
  \citenamefont {Kino},\ and\ \citenamefont {Kamimura}}]{Hiyama:2003cu}%
  \BibitemOpen
  \bibfield  {author} {\bibinfo {author} {\bibfnamefont {E.}~\bibnamefont
  {Hiyama}}, \bibinfo {author} {\bibfnamefont {Y.}~\bibnamefont {Kino}}, \ and\
  \bibinfo {author} {\bibfnamefont {M.}~\bibnamefont {Kamimura}},\ }\href
  {\doibase 10.1016/S0146-6410(03)90015-9} {\bibfield  {journal} {\bibinfo
  {journal} {Prog. Part. Nucl. Phys.}\ }\textbf {\bibinfo {volume} {51}},\
  \bibinfo {pages} {223} (\bibinfo {year} {2003})}\BibitemShut {NoStop}%
\bibitem [{\citenamefont {Tang}\ \emph {et~al.}(1978)\citenamefont {Tang},
  \citenamefont {Lemere},\ and\ \citenamefont {Thompson}}]{Tang:1978zz}%
  \BibitemOpen
  \bibfield  {author} {\bibinfo {author} {\bibfnamefont {Y.}~\bibnamefont
  {Tang}}, \bibinfo {author} {\bibfnamefont {M.}~\bibnamefont {Lemere}}, \ and\
  \bibinfo {author} {\bibfnamefont {D.}~\bibnamefont {Thompson}},\ }\href
  {\doibase 10.1016/0370-1573(78)90175-8} {\bibfield  {journal} {\bibinfo
  {journal} {Phys. Rept.}\ }\textbf {\bibinfo {volume} {47}},\ \bibinfo {pages}
  {167} (\bibinfo {year} {1978})}\BibitemShut {NoStop}%
\bibitem [{\citenamefont {Le~Yaouanc}\ \emph {et~al.}(1973)\citenamefont
  {Le~Yaouanc}, \citenamefont {Oliver}, \citenamefont {Pene},\ and\
  \citenamefont {Raynal}}]{LeYaouanc:1972ae}%
  \BibitemOpen
  \bibfield  {author} {\bibinfo {author} {\bibfnamefont {A.}~\bibnamefont
  {Le~Yaouanc}}, \bibinfo {author} {\bibfnamefont {L.}~\bibnamefont {Oliver}},
  \bibinfo {author} {\bibfnamefont {O.}~\bibnamefont {Pene}}, \ and\ \bibinfo
  {author} {\bibfnamefont {J.}~\bibnamefont {Raynal}},\ }\href {\doibase
  10.1103/PhysRevD.8.2223} {\bibfield  {journal} {\bibinfo  {journal} {Phys.
  Rev.}\ }\textbf {\bibinfo {volume} {D8}},\ \bibinfo {pages} {2223} (\bibinfo
  {year} {1973})}\BibitemShut {NoStop}%
\bibitem [{\citenamefont {Bonnaz}\ and\ \citenamefont
  {Silvestre-Brac}(1999)}]{Bonnaz:1999zj}%
  \BibitemOpen
  \bibfield  {author} {\bibinfo {author} {\bibfnamefont {R.}~\bibnamefont
  {Bonnaz}}\ and\ \bibinfo {author} {\bibfnamefont {B.}~\bibnamefont
  {Silvestre-Brac}},\ }\href {\doibase 10.1007/s006010050128} {\bibfield
  {journal} {\bibinfo  {journal} {Few Body Syst.}\ }\textbf {\bibinfo {volume}
  {27}},\ \bibinfo {pages} {163} (\bibinfo {year} {1999})}\BibitemShut
  {NoStop}%
\bibitem [{\citenamefont {Broyden}(1965)}]{Broyden:1965}%
  \BibitemOpen
  \bibfield  {author} {\bibinfo {author} {\bibfnamefont {C.~G.}\ \bibnamefont
  {Broyden}},\ }\href {\doibase 10.1140/epja/i2010-10929-7} {\bibfield
  {journal} {\bibinfo  {journal} {Math. Comp.}\ }\textbf {\bibinfo {volume}
  {19}},\ \bibinfo {pages} {577} (\bibinfo {year} {1965})}\BibitemShut
  {NoStop}%
\bibitem [{\citenamefont {Ortega}\ \emph {et~al.}(2013)\citenamefont {Ortega},
  \citenamefont {Entem},\ and\ \citenamefont {Fernandez}}]{Ortega:2012rs}%
  \BibitemOpen
  \bibfield  {author} {\bibinfo {author} {\bibfnamefont {P.}~\bibnamefont
  {Ortega}}, \bibinfo {author} {\bibfnamefont {D.}~\bibnamefont {Entem}}, \
  and\ \bibinfo {author} {\bibfnamefont {F.}~\bibnamefont {Fernandez}},\ }\href
  {\doibase 10.1088/0954-3899/40/6/065107} {\bibfield  {journal} {\bibinfo
  {journal} {J. Phys.}\ }\textbf {\bibinfo {volume} {G40}},\ \bibinfo {pages}
  {065107} (\bibinfo {year} {2013})}\BibitemShut {NoStop}%
\bibitem [{\citenamefont {Olive}\ \emph {et~al.}(2014)\citenamefont {Olive}
  \emph {et~al.}}]{PDG2014}%
  \BibitemOpen
  \bibfield  {author} {\bibinfo {author} {\bibfnamefont {K.}~\bibnamefont
  {Olive}} \emph {et~al.} (\bibinfo {collaboration} {Particle Data Group}),\
  }\href {\doibase 10.1088/1674-1137/38/9/090001} {\bibfield  {journal}
  {\bibinfo  {journal} {Chin. Phys.}\ }\textbf {\bibinfo {volume} {C38}},\
  \bibinfo {pages} {090001} (\bibinfo {year} {2014})}\BibitemShut {NoStop}%
\bibitem [{\citenamefont {del Amo~Sanchez}\ \emph {et~al.}(2010)\citenamefont
  {del Amo~Sanchez} \emph {et~al.}}]{PhysRevD.82.011101}%
  \BibitemOpen
  \bibfield  {author} {\bibinfo {author} {\bibfnamefont {P.}~\bibnamefont {del
  Amo~Sanchez}} \emph {et~al.} (\bibinfo {collaboration} {The BABAR
  Collaboration}),\ }\href {\doibase 10.1103/PhysRevD.82.011101} {\bibfield
  {journal} {\bibinfo  {journal} {Phys. Rev. D}\ }\textbf {\bibinfo {volume}
  {82}},\ \bibinfo {pages} {011101} (\bibinfo {year} {2010})}\BibitemShut
  {NoStop}%
\bibitem [{\citenamefont {Aaij}\ \emph {et~al.}(2014)\citenamefont {Aaij} \emph
  {et~al.}}]{Aaij2014665}%
  \BibitemOpen
  \bibfield  {author} {\bibinfo {author} {\bibfnamefont {R.}~\bibnamefont
  {Aaij}} \emph {et~al.},\ }\href
  {http://www.sciencedirect.com/science/article/pii/S0550321314001941}
  {\bibfield  {journal} {\bibinfo  {journal} {Nuclear Physics B}\ }\textbf
  {\bibinfo {volume} {886}},\ \bibinfo {pages} {665 } (\bibinfo {year}
  {2014})}\BibitemShut {NoStop}%
\bibitem [{\citenamefont {Swanson}(2004)}]{Swanson2004197}%
  \BibitemOpen
  \bibfield  {author} {\bibinfo {author} {\bibfnamefont {E.~S.}\ \bibnamefont
  {Swanson}},\ }\href
  {http://www.sciencedirect.com/science/article/pii/S0370269304011463}
  {\bibfield  {journal} {\bibinfo  {journal} {Physics Letters B}\ }\textbf
  {\bibinfo {volume} {598}},\ \bibinfo {pages} {197 } (\bibinfo {year}
  {2004})}\BibitemShut {NoStop}%
\bibitem [{\citenamefont {Takizawa}\ \emph {et~al.}(2014)\citenamefont
  {Takizawa}, \citenamefont {Takeuchi},\ and\ \citenamefont
  {Shimizu}}]{Takizawa:2014nma}%
  \BibitemOpen
  \bibfield  {author} {\bibinfo {author} {\bibfnamefont {M.}~\bibnamefont
  {Takizawa}}, \bibinfo {author} {\bibfnamefont {S.}~\bibnamefont {Takeuchi}},
  \ and\ \bibinfo {author} {\bibfnamefont {K.}~\bibnamefont {Shimizu}},\
  }\bibfield  {booktitle} {\emph {\bibinfo {booktitle} {{Proceedings, 22nd
  European Conference on Few-Body Problems in Physics (EFB22)}}},\ }\href
  {\doibase 10.1007/s00601-014-0830-6} {\bibfield  {journal} {\bibinfo
  {journal} {Few Body Syst.}\ }\textbf {\bibinfo {volume} {55}},\ \bibinfo
  {pages} {779} (\bibinfo {year} {2014})}\BibitemShut {NoStop}%
\end{thebibliography}%

\end{document}